\documentclass[journal]{IEEEtran}
\usepackage{cite}
\usepackage{mathtools,amsmath,amssymb,amsfonts}
\usepackage{cuted}
\usepackage{algorithmic}
\usepackage{graphicx}
\usepackage{textcomp}
\usepackage{xcolor}
\usepackage{comment}
\usepackage{xparse}
\usepackage{booktabs}

\def\BibTeX{{\rm B\kern-.05em{\sc i\kern-.025em b}\kern-.08em
    T\kern-.1667em\lower.7ex\hbox{E}\kern-.125emX}}
\usepackage{color}

\newcommand{\dmin}{\ensuremath{d_{\mathrm{min}}}}%
\begin{document}


\title{Delay Optimization of Conventional Non-Coherent Differential CPM Detection\\
}


\author{\IEEEauthorblockN{Anouar Jerbi, Karine Amis, Frédéric Guilloud, and Tarik Benaddi}\\
\thanks{This work has been financially supported by Thales Alenia Space and Brest Métropole.\\
Anouar Jerbi is with the Mathematical and Electrical Engineering (MEE) Department, IMT Atlantique, Lab-STICC, UMR CNRS 6285, 29238 Brest, France. He is also with TéSA, 31000 Toulouse, France. (e-mail: anouar.jerbi@imt-atlantique.fr).\\
Karine Amis and Frédéric Guilloud are with the Mathematical and Electrical Engineering (MEE) Department, IMT Atlantique, Lab-STICC, UMR CNRS 6285, 29238 Brest, France (e-mail: firstname.lastname@imt-atlantique.fr).
Tarik Benaddi is with Thales Alenia Space, 31000 Toulouse, France (e-mail : tarik.benaddi@thalesaleniaspace.com)}
}
\maketitle
\begin{abstract}
The conventional non-coherent differential detection of continuous phase modulation (CPM) is quite robust to channel impairments such as phase and Doppler shifts. Its implementation is on top of that simple. 
It consists in multiplying the received baseband signal by its conjugate and delayed version of one symbol period. However it suffers from a signal-to-noise ratio gap compared to the optimum coherent detection.
In this paper, we improve the error rate performance of the conventional differential detection by using a delay higher than one symbol period. We derive the trellis description as well as the branch and cumulative metric that take into account a delay of $K$ symbol periods. We then derive an optimization criterion of $K$ based on the minimum Euclidean distance between two differential signals. We finally determine optimized delays for some popular CPM formats whose values are confirmed by error rate simulations.
\end{abstract}


\section{Introduction}
\IEEEPARstart{C}{ontinuous Phase Modulations} (CPM) are a class of non-linear constant-envelope modulations  with limited spectral occupancy. The constant envelope is interesting when the channel includes a strong non-linearity like e.g. in satellite communications.
Moreover, non-coherent CPM detection enables to face the possible phase distortion introduced by the channel, by doing without any phase synchronisation. Combined with the energy efficiency of CPM, these properties make this kind of waveform a good candidate for Internet of Things (IoT) \cite{PKP20,MHN22}.
Non-coherent CPM detectors can be grouped into two families depending on the criterion they are based on. The first one is derived from the generalized maximum-likelihood criterion \cite{HL68,DPT18,JAG22Gretsi}  and only requires the knowledge of the phase distribution. Algorithms proposed either in \cite{SD93} or in \cite{CR99CPM} with an uniformly-distributed phase assumption belong to it. The second one preprocesses the received signal to neutralize the phase contribution making possible the application  of the maximum-likelihood criterion for coherent detection on the resulting signal. 

Numerous papers based on differentially-preprocessed signals can be found in the state-of-the-art. The common feature is the use of the product of the received baseband signal and a conjugate time-delayed version of it, yielding a signal that we will refer to as differential signal in the remaining of the paper. Different algorithms are proposed and apply either on time-discrete differential signals (see e.g. \cite{K89}) or on time-continuous differential signals (see e.g. \cite{MM90,SW83,SS86,MF93}). In \cite{MM90} applied to tamed frequency modulation (TFM) and in \cite{MF93} extended to CPM signals, a detection metric is defined from a set of multiple differential signal versions of the original one (differing from the delay value). Simulations are used to compare different set definitions (with a maximum delay equal to three symbol periods) for the TFM and the Gaussian minimum shift keying (GMSK). However, the most used representative of the second class is the conventional differential detection as defined in \cite{SW83,SS86} and applied with one symbol period as the delay value. The latter is on top of that insensitive to Doppler shifts, which motivated us to focus on this kind of receiver given our interest for SAT-IoT. Its main drawback is the signal to noise ratio (SNR) gap as compared to the optimum coherent detection. In this paper, for the purpose of reducing the SNR gap, we propose to modify the differential detection as defined in \cite{SW83}. Our contributions are threefold, (i) the extension of the usual differential detection algorithm to consider a delay higher than one symbol period (including the definition of the phase trellis and of the branch and cumulative metric), (ii) the proposition of an optimization criterion of the delay value based on the minimum Euclidean distance between two differential signals and, (iii) the optimized delay values for different CPM formats (modulation index, phase shaping pulse length, frequency pulse).
The remainder of this article is organized as follows: in Section~\ref{sec:notations}, the system model is presented and the notations are introduced. In Section~\ref{sec:algo}, the differential detection using a delay of $K$ symbol periods is exposed, followed by the optimization criterion of $K$ in Section \ref{sec:distance_min}. The simulations and the resulting tables for different CPM formats are presented in Section~\ref{sec:simus}. A conclusion is drawn in Section~\ref{sec:conclusion} which ends the paper.

\section{System Model and Notations} \label{sec:notations}
We consider a sequence of $N$ independent and identically distributed (i.i.d.) information symbols $\mathbf{a}=\{a_i\}_{0\leq i\leq N-1}$ to be transmitted. Given $M$ an even positive integer, $a_i$ takes on values in the $M$-ary alphabet $\mathcal{M}=\{\pm 1,\pm 3,..\pm(M-1)\}$ with equal probabilities. The complex envelope of the CPM-modulated signal is given by:\\
\begin{equation}
s(t,\mathbf{a})=\sqrt{\frac{E_s}{T_s}}\mathrm{e}^{j\theta(t,\mathbf{a})},
\end{equation}
where $E_s$ is the average symbol energy, $T_s$ is the symbol period and $\theta (t,\mathbf{a})$ is the signal phase which depends on the information symbols.
It is defined by:
\begin{equation}\label{phase_eq}
\theta(t,\mathbf{a})=2\pi h\displaystyle\sum_{i=0}^{N-1} a_i q(t-iT_s),
\end{equation}
where $h$ is the modulation index and  $q(t)$ is the phase shaping pulse whose expression is $q(t)= \int_{- \infty}^{t} g(u)du$ with $g(t)$ the frequency waveform. In practice, $g(t)$ has a finite duration $LT_s$ and it must satisfy the following conditions:\\
\begin{equation} \label{eq:FreqPulseProperty}
\begin{cases}
    g(t)=g(LT_s-t), & 0\leq t < LT_s\\
    \int_{-\infty}^t g(\tau)\mathrm{d}\tau = q(LT_s) = \frac{1}{2},  & \forall t \geq LT_s
  \end{cases}
\end{equation} 
We assume that the modulated signal is transmitted over a Gaussian channel. The equivalent baseband received signal, denoted by $r(t)$, is given by:
\begin{equation}
r(t) = s(t,\mathbf{a})e^{j\psi} + n(t),
\end{equation}
where $\psi$ is an arbitrary phase introduced by the channel supposed to be uniformly distributed in $[0,2\pi)$ and $n(t)$ is the realization of a zero-mean wide-sense stationary complex circularly symmetric Gaussian noise, independent of the signal, and with double-sided PSD $\frac{N_0}{2}$.
\section{Differential detection of CPM} \label{sec:algo}
\subsection{$K$-delay based differential receiver}
At the receiver side, a differential signal denoted by $R_K(t)$ is generated using the received signal $r(t)$ and its $K-$symbol periods delayed version $r(t-KT_s)$ and is expressed as the sum of two signals:
\begin{eqnarray}
R_{K}(t) = \frac{1}{2} r(t)r^*(t-KT_s)
 =  S_K(t,\mathbf{a})+N_K(t),
\end{eqnarray}
where the first term does not include any noise part: 
\begin{eqnarray}
S_K(t,\mathbf{a}) \frac{1}{2} s(t,\mathbf{a})s^*(t-KT_s,\mathbf{a})
=\frac{E_s}{2T_s}\mathrm{e}^{j\Theta_K(t,\mathbf{a})} \label{eq:ThetaK}
\end{eqnarray}
with $\Theta_K(t,\mathbf{a})=\theta(t,\mathbf{a}) - \theta(t-KT_s,\mathbf{a})$.

The second term which includes noise-dependent component $N_K(t)$ is decomposed as $N_K(t) = U_K(t) + W_K(t)$ with
\begin{eqnarray}
U_K(t) &=& \frac{1}{2} \left( s(t,\mathbf{a})n^*(t-KT_s) + n(t)s^*(t-KT_s,\mathbf{a})\right), \nonumber\\
W_K(t) &=& \frac{1}{2} \left(n(t)n^*(t-KT_s)\right).
\end{eqnarray}
The noise component autocorrelation can be written as:
\begin{eqnarray}
    E[N_K(t)N^*_K(t-\tau)] 
    &=& \frac{1}{4} (N_0^2 + 2 A^2 N_0)\delta(\tau)
\end{eqnarray}
with $A^2=|s(t,\mathbf{a})|= \sqrt{\frac{E_s}{T_s}}$ and $\delta(t)$ the  delta function. The noise process $N_K(t)$ is wide-sense stationary with zero mean and constant PSD equal to $\frac{1}{4} (N_0^2 + 2 A^2 N_0)$. From now on, it will be assumed to follow a Gaussian distribution as in \cite{MF93}.

\subsection{Phase trellis description\label{sec:trellis}}


Let $t~=~\tau ~+ ~nT_s$, with $0\leq \tau < T_s$.
Taking into account the properties of the frequency pulse given in \eqref{eq:FreqPulseProperty}, the phase introduced in \eqref{eq:ThetaK} can be decomposed as the sum of a time-independent term and a time-dependent term:
\begin{eqnarray}
\Theta_{K}(\tau + nT_s,\mathbf{a}) = \phi_n+2\pi h a_n q(\tau) +\varphi_n(\tau),
\label{phase_diff_eq}
\end{eqnarray}
with $\phi_n=\pi h\displaystyle\sum_{i=0}^{K-1} a_{n-L-i}$
and 
\begin{equation}
\varphi_n(\tau)~=~2\pi h \left(\displaystyle\sum_{i=1}^{L-1} (a_{n-i} - a_{n-K-i}) q(\tau + iT_s) - a_{n-K} q(\tau)\right).
\end{equation}
$\varphi_n(\tau)$ represents the time-dependent contribution which corresponds to the last $L$ memory symbols of both the signal and its delayed version. The term $\phi_n$ represents the time-independent part.
$\varphi_n(\tau)$ and $\phi_n$ are completely determined by the set of symbols $(a_{n-i})_{1\leq i \leq L+K-1}$. 
We can thus define the state $\Sigma_n = [a_{n-L-K+1},...,a_{n-1}]$ of the CPM encoder. Note that there are $M^{K+L-1}$ different possible states. The symbol modulation corresponds to a path representing $\Theta_{K}$ in the trellis made of states $\Sigma_n$.

\subsection{Maximum likelihood (ML)-based detection}
The ML criterion is applied to detect the information symbols from $R_K(t)$. Given the constant amplitude property of CPM, it consists in maximizing the correlation between $R_K(t)$ and all possible realizations of $S_K(t)$. The inner product between $R_K(t)$ and $S_K(t)$, denoted by $\Gamma_N\tilde{\mathbf{a}}) $, is defined as
\begin{equation}
\Gamma_N(\tilde{\mathbf{a}}) = \int_{-\infty}^{NT_s} R_K(t)S_K(t,\tilde{\mathbf{a}})dt,
\end{equation}
which can be recursively computed:
\begin{equation} \label{eq:BranchMetric}
\Gamma_n(\tilde{\mathbf{a}}) = \Gamma_{n-1}(\tilde{\mathbf{a}})+\Lambda_n(\tilde{\mathbf{a}})
\end{equation}
with
\begin{equation}
\Lambda_n(\tilde{\mathbf{a}}) = \int_{nT_s}^{(n+1)T_s} R_K(t)S_K(t,\tilde{\mathbf{a}})dt.
\end{equation}
The Viterbi algorithm is applied on the trellis. At the $n-$th section, it computes for each state the maximum cumulative metric \eqref{eq:BranchMetric} among all the paths arriving at this state.

\section{Proposed delay optimization based on the minimum Euclidean distance criterion}\label{sec:distance_min}

In this Section, we aim at tuning $K$ to improve the detection error probability. Let us consider the following error event when $s(t,\mathbf{a})$ is transmitted and $s(t,\tilde{\mathbf{a}})$ is detected and $\mathbf{a}\neq \tilde{\mathbf{a}}$. Given the ML-based detection criterion and the independence between $R_K$ and $N_K$, it means that:
\begin{equation}\label{error_eq}
    \int_0^{NT_s} | R_K(t) - S_K(t,\tilde{\mathbf{a}}) |^2 dt \leq \int_0^{NT_s} | R_K(t) - S_K(t,\mathbf{a}) |^2 dt
\end{equation}
which can be reformulated as:
\begin{equation}
    Z_K \geq \frac{1}{2} \Delta^2_K(\mathbf{a},\tilde{\mathbf{a}}),
\end{equation}
where $Z_K = \int_0^{NT_s} \mathrm{Re}\left[ \left( S_K(t,\mathbf{a}) - S_K(t,\tilde{\mathbf{a}}) \right)N^{*}_K(t) \right] dt$. $\Delta_K(\mathbf{a},\tilde{\mathbf{a}}) = \sqrt{\int_0^{NT_s} \left| S_K(t,\mathbf{a}) - S_K(t,\tilde{\mathbf{a}}) \right|^2 dt}$ is the Euclidean distance between the two differential signals $S_K(t,\mathbf{a})$ and $S_K(t,\tilde{\mathbf{a}})$  corresponding to the symbol sequences $\mathbf{a}$ and $\tilde{\mathbf{a}}$. 
$Z_K$ has zero mean. Assuming that $Z_K$ is Gaussian, the probability of an error event is given by
\begin{equation}
P_e(\mathbf{a};\tilde{\mathbf{a}}) 
= Q \left( \sqrt{\frac{4 \varepsilon_b}{N_0^2 + 2 A^2 N_0}d^2_K(\mathbf{a},\tilde{\mathbf{a}})} \right)
\end{equation}
where $Q$ is the $Q-$function and where  $d_K(\mathbf{a},\tilde{\mathbf{a}}) = \frac{\Delta_K(\mathbf{a},\tilde{\mathbf{a}})}{\sqrt{2\varepsilon_b}}$ is the normalized Euclidean distance, $\varepsilon_b$ denoting the average energy per information bit in the differential symbol sequence.
Proceeding as in \cite{AAS86} (Chapter 2, Paragraph 2.1.2), a union bound on the probability of error is obtained  at reasonably high SNR. The error probability is thus approximated by 
\begin{equation}\label{performance_eq}
P_e \propto Q\left( \sqrt{\frac{4 \varepsilon_b}{N_0^2 + 2 A^2 N_0}\dmin^2(K)}\right)
\end{equation}
where $\dmin(K)$ denotes the  minimum Euclidean distance between two differential signals:
\begin{equation}
\dmin^2(K) = \min_{\substack{\mathbf{a}, \mathbf{\tilde{a}} \\ a_0 \neq \tilde{a_0}}} \left( d^2_K(\mathbf{a},\tilde{\mathbf{a}}) \right)
\end{equation}
By applying the same reasoning as in \cite{AAS86}, we obtain another formulation of the Euclidean distance:
\begin{equation} \label{eq:EuclideanDistance}
    d^2_K(\mathbf{a},\tilde{\mathbf{a}}) = \frac{\log_2(M)}{T_s}\int_0^{NT_s} [1-\cos\left(\Theta_K(t,\mathbf{e})\right)] dt
\end{equation}
where $\mathbf{e}=\mathbf{a}-\tilde{\mathbf{a}}$ is the so-called difference symbol sequence.

Finding the minimum Euclidean distance is done by searching over all possible pairs of sequences $\mathbf{a}$ and $\tilde{\mathbf{a}}$. In practice, these pairs are those whose respective paths on a phase tree diverge at time 0 and merge again as soon as possible.
Proceeding as in \cite{AAS86}, the phase difference tree is a good method to determine the difference symbol sequences to be considered and the corresponding pairs of symbol sequences.

For each value of the delay, a corresponding value of the minimum Euclidean distance $\dmin$ is obtained. Since we are looking for minimizing the error probability, the best choice of the delay is the value that yields the highest $\dmin$.


\section{Numerical results} \label{sec:simus}
In this Section, we study different CPM formats.
In all cases, we consider short frames communications (symbol sequence length $N=120$) and a symbol duration $T_s = 10^{-4}$ s.
\subsection{Influence of $K$ on the detection performance}
\begin{figure}[t]
  \centering
   \includegraphics[scale=0.42]{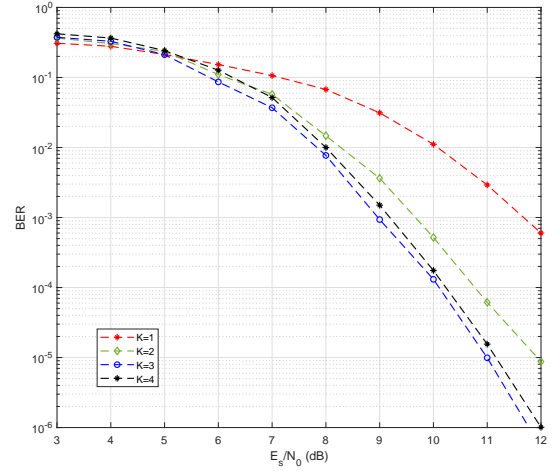}
   \caption{BER of differential detection for the CPM scheme 3REC with modulation modulation index $h=0.75$ for different values of delay $K$}
   \label{fig:3REC34}
\end{figure}

We first illustrate the influence of $K$ on the detection performance. We consider a CPM format with rectangular frequency pulse, $L=3$ and $h=0.75$. The delay $K$ takes on values in \{1,2,3,4\}. The Bit Error Rate (BER) is plotted as a function of $E_b/N_0$ in Fig.~\ref{fig:3REC34}.
We observe that $K=3$ is  the delay that yields the best BER. A gain of 3 dB is obtained compared to the receiver with $K = 1$ and almost 1 dB compared to the receiver with $K = 2$ while the receiver with $K=4$ exhibits  a slight degradation of performance.

\subsection{Optimization of $K$ from the minimum Euclidean distance criterion}
The optimization over $K$ of the Euclidean distance in \eqref{eq:EuclideanDistance} is run by Monte-Carlo simulations by considering several possible pairs of sequences yielding  different possible realizations of $\mathbf{e}$ for several CPM families and several parameters.  The optimized value of $K$ is provided in Tables~\ref{tab:K_optim_values_RC},\ref{tab:K_optim_values_REC},\ref{tab:K_optim_values_GFSK} for Raised Cosine, rectangular and Gaussian frequency shaping pulses respectively, for several modulation indices and several pulse lengths. The coherence of these optimized delays has also been checked by BER simulations, but the curves are omitted in this article due to space limitations. Note that when several values of $K$ provide the best error rate, then the displayed value  is simply the lowest one to reduce the complexity of the decoder.
\begin{table}[t]
\centering
\caption{Optimized values of $K$ for RC CPM 
}
\label{tab:K_optim_values_RC}
\begin{tabular}[t]{@{}cccc@{}}
\toprule
 \multicolumn{1}{c}{Freq. pulse} & \multicolumn{3}{c}{Modulation index} \\
 \multicolumn{1}{c}{length $L$} &$h=1/3$ &$h=1/2$ &$h=3/4$\\
\midrule 
$1$&$K=2$&$K=2$&$K=3$\\
$3$&$K=3$&$K=3$&$K=3$\\
$5$&$K=4$&$K=4$&$K=4$\\
\bottomrule
\end{tabular}
\end{table}

\begin{table}[t]
\centering
\caption{Optimized values of $K$ for REC CPM 
}
\label{tab:K_optim_values_REC}
\begin{tabular}[t]{@{}cccc@{}}
\toprule
 \multicolumn{1}{c}{Freq. pulse} & \multicolumn{3}{c}{Modulation index} \\
 \multicolumn{1}{c}{length $L$} &$h=1/3$ &$h=1/2$ &$h=3/4$\\
\midrule 
$1$&$K=2$&$K=2$&$K=4$\\
$3$&$K=4$&$K=4$&$K=3$\\
$5$&$K=5$&$K=5$&$K=5$\\
\bottomrule
\end{tabular}
\end{table}

\begin{table}[t]
\centering
\caption{Optimized values of $K$ for GFSK $(BT=0.3)$ 
}
\label{tab:K_optim_values_GFSK}
\begin{tabular}[t]{@{}cccc@{}}
\toprule
 \multicolumn{1}{c}{Freq. pulse} & \multicolumn{3}{c}{Modulation index} \\
 \multicolumn{1}{c}{length $L$} &$h=1/3$ &$h=1/2$ &$h=3/4$\\
\midrule 
$3$&$K=3$&$K=3$&$K=4$\\
$5$&$K=3$&$K=3$&$K=4$\\
\bottomrule
\end{tabular}
\end{table}

\subsection{Comparison with some state-of-the-art receivers}
\begin{figure}[t]
  \centering
   \includegraphics[scale=0.42]{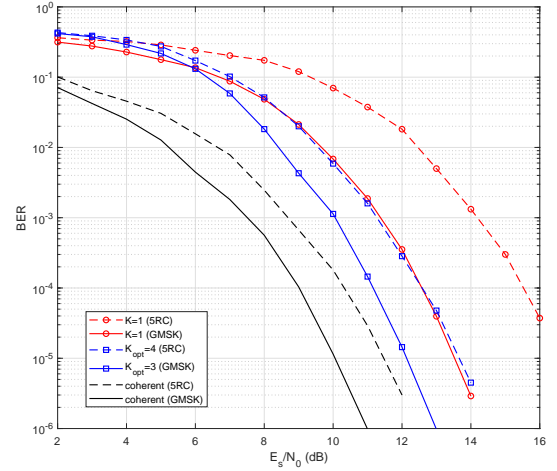}
   \caption{BER comparison between coherent and differential detection for two CPM schemes: GMSK with $BT=0.3$, and 5RC with $h=0.5$}
   \label{fig:BER_GMSK_5RC}
\end{figure}

In Fig.~\ref{fig:BER_GMSK_5RC}, we show a comparison between the optimized differential receiver, the $K=1$ differential receiver, and also the coherent receiver. This comparison is performed for 2 different CPM families: GMSK with BT=0.3, and 5RC with $h=0.5$. For GMSK, there is almost $4$dB  between the coherent BER and the $K=1$ differential detection. Using the optimized $K=3$ reduces this gap by almost $2$dB.
For the 5RC CPM, using the optimized $K=4$ delay reduces the gap to coherent BER from around $6$dB down to $2$dB. Note that the curves for the coherent and the optimized differential receivers are quasi-parallel which means that the diversity gain is almost the same and the difference between the two is mainly in the noise variance which is higher for the differential receiver.

\subsection{Comparison in presence of Doppler shift}

\begin{figure}[t]
  \centering
   \includegraphics[scale=0.42]{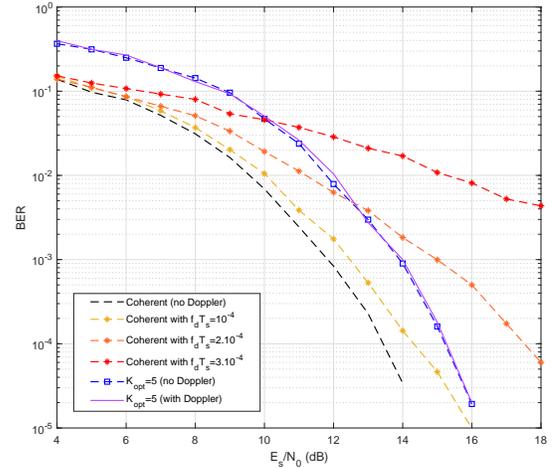}
   \caption{BER comparison between coherent and differential detection for 5REC with $h=0.5$ in the presence of a Doppler shift}
   \label{fig:BER_5REC_12_doppler}
\end{figure}

The differential detector is especially interesting in applications where the Doppler shift affects the communication. In Fig.~\ref{fig:BER_5REC_12_doppler}, a performance comparison between the differential detector and the coherent one in terms of BER is illustrated for the rectangular pulse with $h=0.5$ and $L=5$ in presence of a small Doppler shift. We see a huge degradation of performances for the coherent detector whereas the differential detector is not affected, whatever the Doppler shift value.

\section{Conclusion} \label{sec:conclusion}

In this article, the increase of the delay used in the conventional non-coherent differential detection of CPM is shown to have an impact on the error rate. We therefore propose an optimization criterion of this delay based on the minimum Euclidean distance between two differential signals. This optimized delay ranges from 2 to 5 symbol periods depending on the considered CPM format. Simulations confirm the choice of the optimized delay value which offers a gain from 2 to 4 dB on the error rate performance compared to a single symbol duration delay. CPM with optimized differential detection may be an alternative candidate waveform in the perspective of limited-power Satellite IoT where Doppler shift is an issue.

\bibliographystyle{IEEEtran}
\bibliography{biblio}

\begin{thebibliography}{10}
\providecommand{\url}[1]{#1}
\csname url@samestyle\endcsname
\providecommand{\newblock}{\relax}
\providecommand{\bibinfo}[2]{#2}
\providecommand{\BIBentrySTDinterwordspacing}{\spaceskip=0pt\relax}
\providecommand{\BIBentryALTinterwordstretchfactor}{4}
\providecommand{\BIBentryALTinterwordspacing}{\spaceskip=\fontdimen2\font plus
\BIBentryALTinterwordstretchfactor\fontdimen3\font minus
  \fontdimen4\font\relax}
\providecommand{\BIBforeignlanguage}[2]{{%
\expandafter\ifx\csname l@#1\endcsname\relax
\typeout{** WARNING: IEEEtran.bst: No hyphenation pattern has been}%
\typeout{** loaded for the language `#1'. Using the pattern for}%
\typeout{** the default language instead.}%
\else
\language=\csname l@#1\endcsname
\fi
#2}}
\providecommand{\BIBdecl}{\relax}
\BIBdecl

\bibitem{PKP20}
A.~G. Perotti, M.~Khormuji, and B.~Popović, ``{Simultaneous Wireless
  Information and Power Transfer by Continuous-Phase Modulation},'' \emph{IEEE
  Communication Letters}, vol.~24, no.~6, pp. 1294--1298, 2020.

\bibitem{MHN22}
A.~Mondal, M.~Hanif, and H.~Nguyen, ``{SSK-ICS LoRa: A LoRa-Based Modulation
  Scheme with Constant Envelope and Enhanced Data Rate},'' \emph{IEEE
  Communication Letters}, pp. 1--4, 2022.

\bibitem{HL68}
H.~L.~V. Trees, \emph{Detection, estimation and modulation theory}.\hskip 1em
  plus 0.5em minus 0.4em\relax New York: John Wiley \& Sons, vol. I, 1968.

\bibitem{DPT18}
C.~Piat-Durozoi, C.~Poulliat, N.~Thomas, M.~Boucheret, G.~Lesthievent, and
  E.~Bouisson, ``{Minimal State Non-Coherent Symbol MAP Detection of
  Continuous-Phase Modulations},'' \emph{IEEE Communication Letters}, vol.~22,
  no.~10, pp. 2008--2011, 2018.

\bibitem{JAG22Gretsi}
A.~Jerbi, K.~Amis, F.~Guilloud, and T.~Benaddi, ``{Détection non-cohérente
  des modulations CPM en présence d'un décalage Doppler},'' submitted for
  presentation at the \textit{GRETSI 2022}.

\bibitem{SD93}
M.~Simon and D.~Divsalar, ``Maximum-likelihood block detection of noncoherent
  continuous phase modulation,'' \emph{IEEE Transactions on Communications},
  vol.~41, no.~1, pp. 90--98, 1993.

\bibitem{CR99CPM}
G.~Colavolpe and R.~Raheli, ``Noncoherent sequence detection of continuous
  phase modulations,'' \emph{IEEE Transactions on Communications}, vol.~47,
  no.~9, pp. 1303--1307, 1999.

\bibitem{K89}
G.~Kawas~Kaleh, ``Differential detection of partial response continuous phase
  modulation with index 0.5,'' in \emph{IEEE 39th Vehicular Technology
  Conference}, 1989, pp. 115--121 vol.1.

\bibitem{MM90}
D.~Makrakis and P.~Mathiopoulos, ``Differential detection of correlative
  encoded continuous phase modulation schemes using decision feedback,'' in
  \emph{IEEE Int. Conf. on Communications}, 1990, pp. 619--625 vol.2.

\bibitem{SW83}
M.~Simon and C.~Wang, ``Differential versus limiter - discriminator detection
  of narrow-band fm,'' \emph{IEEE Transactions on Communications}, vol.~31,
  no.~11, pp. 1227--1234, 1983.

\bibitem{SS86}
N.~Svensson and C.-E. Sundberg, ``Performance evaluation of differential and
  discriminator detection of continuous phase modulation,'' \emph{IEEE
  Transactions on Vehicular Technology}, vol.~35, no.~3, pp. 106--117, 1986.

\bibitem{MF93}
D.~Makrakis and K.~Feher, ``Multiple differential detection of continuous phase
  modulation signals,'' \emph{IEEE Transactions on Vehicular Technology},
  vol.~42, no.~2, pp. 186--196, 1993.

\bibitem{AAS86}
J.~B. Anderson, T.~Aulin, and C.~Sundberg, \emph{Digital Phase Modulation.
  Applications of Communications Theory}.\hskip 1em plus 0.5em minus
  0.4em\relax Springer US, 1986.

\end{thebibliography}

\end{document}